\documentclass[twocolumn,showpacs,preprintnumbers,amsmath,amssymb]{revtex4}
\usepackage{graphicx}% Include figure files
\usepackage{dcolumn}% Align table columns on decimal point
\usepackage{bm}% bold math
\usepackage{psfig}
\begin{document}
%\preprint{FIS-UI-TH-XX-02}

\title{Statistical properties of the Jakarta and Kuala Lumpur 
stock exchange indices before and after crash}
\author{T. Mart} 
\affiliation{Jurusan Fisika, FMIPA, Universitas Indonesia, Depok 16424, 
  Indonesia}

\date{\today}

\begin{abstract}
 Using the tools developed for statistical physics, 
 we simultaneously analyze statistical properties of the Jakarta and 
 Kuala Lumpur Stock Exchange indices. In spite of the small number of 
 data used in the analysis, the result shows the universal behavior of complex 
 systems previously found in the leading stock indices. 
 We also analyze their features before and after the financial 
 crisis. We found that  after the crisis both stocks do not show  
 a same statistical behavior. The impact of currency controls
 is observed in the distribution of index returns.
\end{abstract}


\maketitle

\section{Introduction}
Recently, there have been considerable efforts to analyze financial data
by means of methods developed for statistical physics \cite{canning1998,%
mills2000,raberto,liu1999,gopi2000,mantegna1995,mantegna1999,plerou1999,%
gopi1999,lillo1999,lillo2000a,johansen1998,bonanno1999,wang2001}. 
Motivated by the
scientific challenge to understand the nature of complex systems, physicists
have started to direct their attentions to a huge, and also growing amount of
economics
data recorded minutes by minutes for decades. Among these interesting data,
the fluctuation of stock exchange indices is of special interest, since
it  might indirectly reflect the economic situation in a certain 
region and some people happily speculate their money on it. Furthermore,
the advancement in computing capabilities has enabled them to
handle a large amount of data, unlike almost 40 years ago when
Mandelbrot investigated approximately 2000 data points of cotton
prices \cite{mandelbrot}.

It is then expected that such studies could explain the nature of
interacting elements in the complex system and, therefore, could help to
forecast economic fluctuations in the future. In other words, these studies
were intended to produce new results in economics, 
which might help us to avoid economic ``earthquakes'' such as what happened in
Indonesia a few years ago \cite{stanley2000}. 

\begin{figure}[!b]
  \begin{center}
    \leavevmode
    \psfig{figure=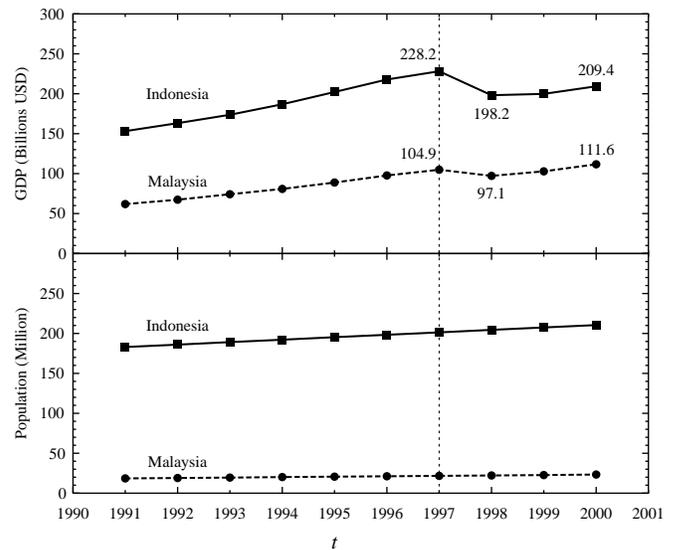,width=90mm}
    \caption{The Indonesian and Malaysian Gross Domestic Products 
        along with their populations. The vertical line indicates the year 
        when the monetary and financial crisis started to occur 
        in the Asia region. In both countries the financial crisis was
        started on July 1997, the impact was naturally recorded as
        the 1998 GDP.
        Data are taken from \protect\cite{doe}.}
   \label{fig:gdp} 
  \end{center}
\end{figure}

Previous studies in this subject so far have focused only on the long-term 
behavior of the leading stock indices. This is understandable, since to
statistically investigate the {\it universal features} in 
economic activities one has to have a large amount of data. Very little
attentions are given to investigate what happen to the stock markets 
in developing countries (e.g., Indonesia) 
as well as what happen to the stock indices 
before, during, and after a financial (or monetary) crash, although 
in the latter the most important ingredients of the 
financial market or economic fluctuation 
could exist \cite{lillo1999,lillo2000a,johansen1998}.

It is the objective of this paper to study the general statistical 
properties of the fluctuation in stock indices in two developing 
countries, as well as their properties before and after a financial crashes.
This study is important in order to investigate the extent of the 
{\it universality} of complex behavior found previously 
in the leading stock indices, such as S\&P 500 and NYSE. 
Such investigation will naturally shed important
information on the variation of the {\it universal constants} in the
scaling behavior of the stock index. With this information at hand 
it is then possible to identify the statistical properties that 
quantify different behaviors in stock markets and 
those which indicate a crash or stable condition.
For this purpose we take two different indices, the Jakarta Stock 
Exchange Index (abbreviated with IHSG, an acronym of {\it Indeks Harga
Saham Gabungan} or composite stock exchange price index) and the 
Kuala Lumpur
Stock Exchange index (KLSE), which belong to different countries. Comparing 
the two indices would be very interesting since both Indonesia and Malaysia 
underwent the same monetary crisis in 1997, which are then followed by 
financial crashes in almost all economic sectors, but with quite different 
economic situations. As has been often discussed, the behavior of two stock
indices could be very different although the two stock markets are situated
in the same region.

Before proceeding to the analysis, it is worth to glance at the 
Indonesian and Malaysian Gross Domestic Products (GDP)  
shown in Fig.\,\ref{fig:gdp}. Before the crisis the growth rate of the 
Indonesian GDP seems to be faster than that of Malaysian GDP. 
The monetary crisis in 1997 has a clear impact on both GDPs. Nevertheless,
the impact on Indonesian GDP is more obvious.
After the crisis, Malaysia was rather successful
to stabilize its economic situation and did not accept the International
Monetary Fund (IMF) prescription. Three years after that Malaysia was 
able to put its GDP slightly above the previous value in 1997.
On the other hand, Indonesia was unable to stop the declination of 
Rupiah against US Dollar and decided to ask the 
IMF to help stabilizing its monetary condition. Other crises, 
which followed the monetary crisis, are blamed as the reasons of this.

\section{General Properties}
\label{general}

\subsection{Time Series and Index Returns}
In our analysis we investigate the daily index returns, which are
defined as
\begin{eqnarray}
  \label{eq:return}
  Z_{\Delta t}(t) &=& \ln Y(t+\Delta t) - \ln Y(t) \nonumber\\
  &=& \ln\, [Y(t+\Delta t)/Y(t)] ,
\end{eqnarray}
with $Y(t)$ indicates the closing index of the stock at day $t$.
Some previous studies \cite{mantegna1995,mantegna1999} are performed 
with the return defined 
as the difference in the index, instead of the difference in the logarithm of
index as given by Eq.\,(\ref{eq:return}). However, in our analysis
we found that the results of calculations by using both 
methods do not differ significantly.

The available data for the IHSG index are the daily closing index data 
recorded from January
1988 to April 2002 which consist of 3526 data points. The KLSE data
contain also the daily closing index 
starting from December 1993 and ending with June 2002, which comprise totally 
2104 data points. The number of data in both indices seems to be the first
obstacle in this analysis, since compared with the previous analysis 
on the S\&P 500 index, e.g. Ref. \cite{gopi1999}, which used 
approximately $5\times 10^6$ data points, the number turns out to be 
extremely small. However, as shown by Mandelbrot in his analysis
on the cotton prices, even with about 2000 data points it is still 
possible to extract a quantitative conclusion from the data.

\begin{figure}[!t]
  \begin{center}
    \leavevmode
    \psfig{figure=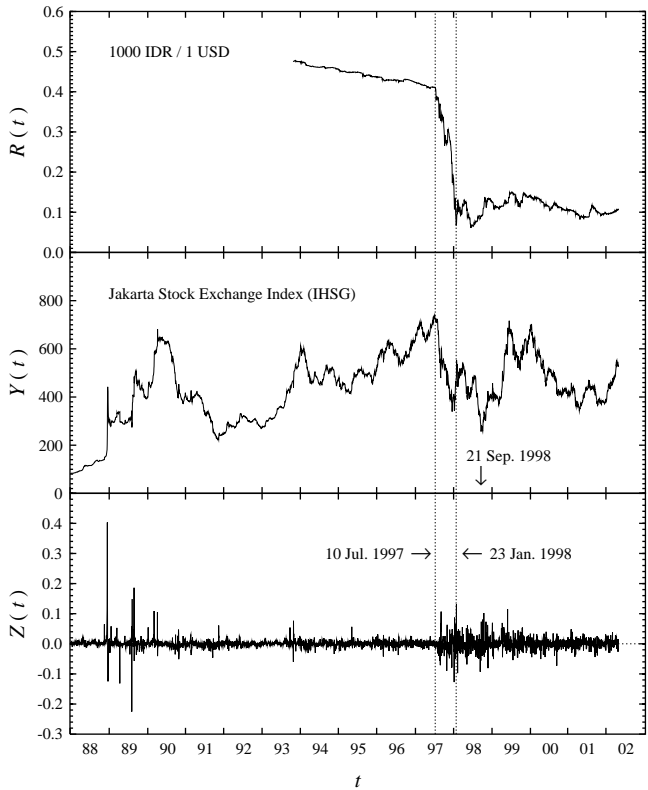,width=90mm}
    \caption{The ratio between 1000 Indonesian Rupiah and 1 US Dollar
        (top), the Jakarta Stock Exchange Index (IHSG) $Y(t)$ (middle) and
        the logarithmic return $Z(t)$ calculated from Eq.\,(\ref{eq:return})
        as a function of time $t$ sampled with 
        $\Delta t=1$ day (bottom). The arrow in the middle panel
        indicates the lowest value position of $Y(t)$ in 1998. 
        Two vertical dashed lines in the figure 
        define the crash period during which the Rupiah exchange rates
        dropped significantly relative to US Dollar. Data are taken from
        \protect\cite{oanda} (IDR/USD exchange rate) and \protect\cite{bej}
        (IHSG stock index).}
   \label{fig:ihsg_return} 
  \end{center}
\end{figure}

Figure \ref{fig:ihsg_return} shows the time series of the IHSG index 
along with the logarithmic returns calculated by using 
Eq.\,(\ref{eq:return}). It is naturally difficult to define when exactly 
the crisis (or crash) started and when it finished. To get around
this difficulty, 
in Fig.\,\ref{fig:ihsg_return} we also display the historical time series
of the exchange rate between 1000 Indonesian Rupiah and 1 US Dollar, since
the economic crash started with the decline of this rate. The
available data do not fill the entire range, nevertheless they are 
sufficient to locate the period of the crisis. It should be noted that
before the monetary crisis the government intervention on this exchange
rate was very strong and as a consequence, although the Rupiah was not 
pegged to US Dollar with a fixed rate, the fluctuation in the exchange 
rate was relatively tiny.

In Fig.\,\ref{fig:ihsg_return}
we indicate the period when Rupiah started to continuously drop 
(10th July 1997) until it reached the minimum point (23rd January 1998).
The IHSG index also dropped significantly during this period. 
However, the index continued to strongly fluctuate and reached another
minimum about one year later due to the unfortunate political situation.
The same phenomenon also happened in the foreign exchange rate, though
with a different scale. In view of this, according to the IHSG index,
the duration of financial crisis could be longer than one year. 

\begin{figure}[!t]
  \begin{center}
    \leavevmode
    \psfig{figure=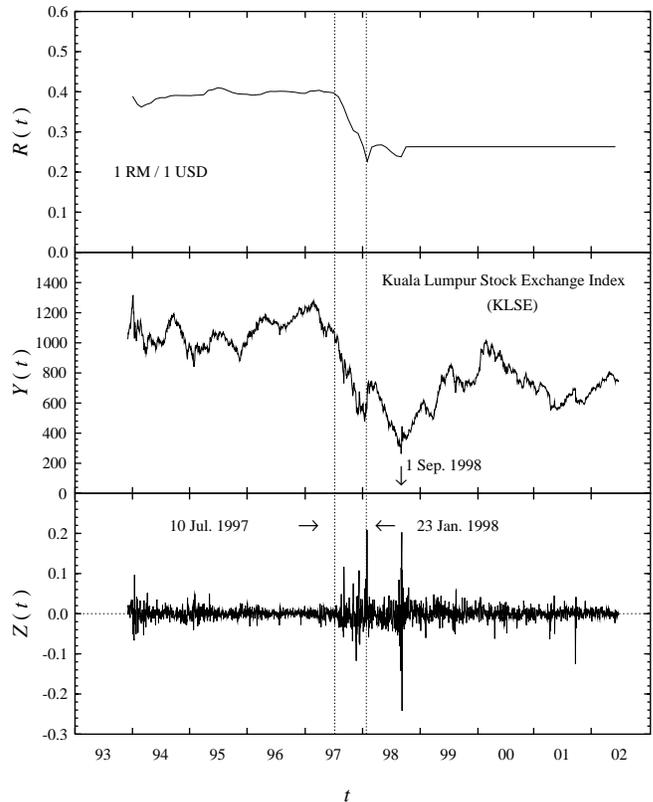,width=90mm}
    \caption{The ratio between 1 Malaysian Ringgit and 1 US Dollar (top), 
        the Kuala Lumpur Stock Exchange Index (IHSG) $Y(t)$ (middle) and
        the logarithmic return $Z(t)$ (bottom) as a function of time
        sampled for $\Delta t=1$ day. On 1st September 1998 Malaysia 
        imposed currency controls, the Ringgit was pegged with US Dollar 
        with a fixed rate of 3.80 Ringgit per Dollar. Data are taken from 
        \protect\cite{oanda} (RM/USD exchange rate) and 
        \protect\cite{yfinance}  (KLSE stock index). 
        Everything else is as in Fig.\,\ref{fig:ihsg_return}.}
   \label{fig:klse_return} 
  \end{center}
\end{figure}

Figure \ref{fig:klse_return} displays the historical time series of the 
KLSE index
along with their logarithmic returns and the ratio between 1 Malaysian
Ringgit and 1 US Dollar. Coincidentally, our definition of the crash period
fits nicely with the latter. As in the case of IHSG, the KLSE index also
significantly dropped during this period. The difference between Indonesian 
and Malaysian foreign exchange rate is, nevertheless, 
apparent here, since on 1st September
1998 Malaysia imposed currency controls, the Ringgit was pegged with US
Dollar with a fixed rate of 3.80 Ringgit per Dollar, while Indonesia was
unable to do that due to the multi-crisis that simultaneously 
occurred in this country.
From Fig.\,\ref{fig:klse_return} it is not clear whether or not 
such policy helped to elevate the KLSE index, since comparing with the
IHSG fluctuation in Fig.\,\ref{fig:ihsg_return} the two indices seem to
be strongly correlated, thus other external factors could be more 
relevant to explain the improvement in the KLSE index. In fact,
Ref.\,\cite{muniandy2001} claims that the policy has led to a misposition
of the Ringgit relative to its realistic exchange rate. 
During 1999 the Ringgit was overvalued since the regional currencies such as
Japan Yen and Singapore Dollar have all depreciated against US Dollar.
As a consequence, Malaysian export became less competitive and 
eventually this policy led to an economic slowdown in Malaysia.

A quick glance to the index fluctuation $Y(t)$ in both figures 
reveals that the fluctuation is more dramatic than that of 
S\&P 500, indicating that in this case the situation is more complex. 
After the crash the magnitude of returns $Z(t)$ is obviously larger 
in both indices, or, in the economics language, the probability to gain 
or to loose becomes larger than before.
In the next section, it will be shown
that both indices are clearly more volatile after the crash.

\begin{figure}[!t]
  \begin{center}
    \leavevmode
    \psfig{figure=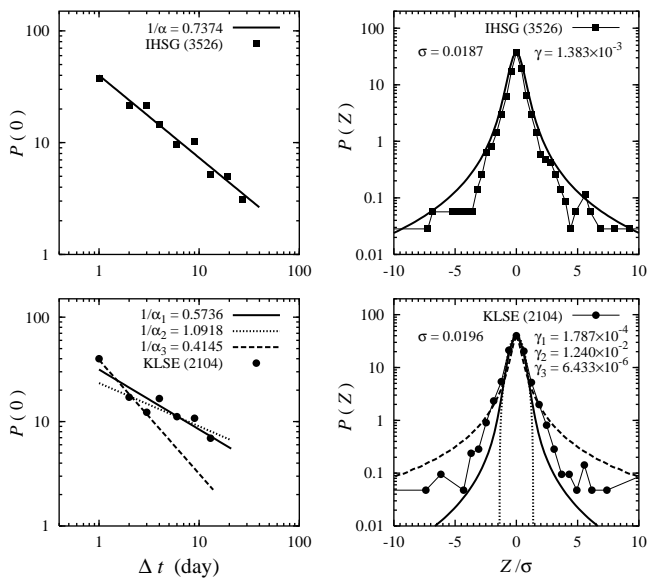,width=90mm}
    \caption{{\it Left panels}: Log-log plots of the IHSG (top) 
      and KLSE (bottom) probability of return
      to the origin as a function of the sampling time $\Delta t$. Solid
      curves show linear regressions to this probability. The dashed curve
      in the case of KLSE shows another possibility if one uses only 
      the first three points in the linear
      regression, whereas the dotted curve is a linear regression by 
      excluding the first point ($\Delta t=1$).
      The slopes of
      these lines, which are equal to negative inverse of the L\'evy
      stable distribution indices, are shown in the left panels.
      {\it Right panels}: Comparison between normalized distribution
      functions for $\Delta t=1$ and normalized L\'evy stable distributions
      (solid and dashed lines) with parameters obtained from the left 
      figures. The
      standard deviation and the number of data are shown in the right
      panels.}
   \label{fig:distxyf0} 
  \end{center}
\end{figure}

\subsection{Scaling the Index Returns}

Following previous studies
\cite{mantegna1995,mantegna1999} we investigate
the probability density function (PDF or $P(Z)$) 
of the return to the origin $P(0)$ in order to 
investigate the scaling behavior of the IHSG and KLSE index returns. 
The advantage of such analysis is obvious, since the number of data included
is relatively small, while the probability is largest
at $Z=0$, thus reducing the statistical inaccuracies. 

Starting with the characteristic function \cite{mantegna1999}
\begin{eqnarray}
  \label{eq:cfunction}
  \varphi(\Delta t) ~=~ e^{-\gamma \Delta t|q|^\alpha} ~,
\end{eqnarray}
the L\'evy stable distribution is given by
\begin{eqnarray}
  \label{eq:levy}
  P (Z) ~=~ \frac{1}{\pi}\, \int_{0}^{\infty}\, 
  e^{-\gamma \Delta t|q|^\alpha} \cos (qZ)\, dq ~.
\end{eqnarray}
From Eq.\,(\ref{eq:levy})
the probability of return to the origin $P(0)$ reads 
\begin{eqnarray}
  \label{eq:pz0}
  P(0) ~=~ \frac{\Gamma (1/\alpha)}{\pi\alpha(\gamma \Delta t)^{1/\alpha}} ~,
\end{eqnarray}
where $\Gamma$ indicates the Gamma function. 

The log-log plots of $P(0)$ as a function of the sampling time $\Delta t$ 
for both IHSG and KLSE indices are shown in the left panels of 
Fig.\,\ref{fig:distxyf0}. Since the number of data points is relatively small  
we limit our analysis only 
up to $\Delta t=19$ days for the IHSG index and $\Delta t=13$ days for KLSE.
In the former (latter) case the number of data points in each set
decreases from 3526 (2104) for $\Delta t=1$ day to the value of 181 for 
$\Delta t=19$ days (160 for $\Delta t=13$ days). These values are
already much smaller compared with previous 
analyses, e.g., for the Hang Seng index \cite{wang2001} one 
ends up with the value of 1481 for $\Delta t=128$ minutes.

The slopes of linear regressions to these plots
equal the negative inverse of the L\'evy stable distribution indices 
$\alpha$. Using this index we calculate the parameter $\gamma$ by means of
Eq.\,(\ref{eq:pz0}) and plot the ``theoretical'' PDF as a function 
of normalized returns $Z/\sigma$, using Eq.\,(\ref{eq:levy}), 
where $\sigma$ the standard deviation of
the distribution, and compare it with the empirical PDF obtained
from data in the right panels of Fig.\,\ref{fig:distxyf0}. 
All relevant parameters obtained 
in this calculation are displayed in Table \ref{tab:properties}.

\begin{table*}[!]
\renewcommand{\arraystretch}{1.2}
\caption{\label{tab:properties} Summary of general statistical properties 
  of the IHSG and KLSE returns $Z(t)$. Linear regression to 
  the probability density function at zero return PDF(0) as a function
  of $\Delta t$ shown in Fig. \ref{fig:distxyf0} 
  yields a line $y=mx+n$, where $m=-1/\alpha$. The standard deviation,
  kurtosis, and skewness are calculated using the formulas given in
  Ref.\,\protect\cite{num_recipes}.}
\begin{ruledtabular}
\begin{tabular}{lcccc}
Properties & IHSG & \multicolumn{3}{c}{KLSE} \\
\cline{3-5}
&&1&2&3\\
\hline
Standard deviation $\sigma$ & 0.0187 & {0.0196}& {0.0196}& {0.0196}\\
Kurtosis & 81.796 & {29.044}& {29.044}& {29.044}\\
Skewness & 3.3987&  {$0.3888$}&  {$0.3888$}&  {$0.3888$}\\
$m$ & $-0.7374$ & $-0.5736$& $-1.0918$& $-0.4145$\\
$n$ &1.6020 & 1.4985& 1.5914& 1.3691\\
$\gamma$ & $1.383\times 10^{-3}$ &
$1.787\times 10^{-4}$& $1.240\times 10^{-2}$& $6.433\times 10^{-6}$\\
$\alpha$ & $1.3561$ & $1.7433$& $0.9159$& $2.4125$\\
$N$ & 3526 & {2104}& {2104}& {2104} \\
\end{tabular}
\end{ruledtabular}
\end{table*}
In the IHSG case we obtain $m=-0.7374$ (corresponds to $\alpha =1.3561$)
 which is slightly 
smaller than that of the S\&P 500 index obtained from the same analysis
\cite{mantegna1999} 
($\alpha =1.40$), or using different approach \cite{gopi1999} (
$\alpha = 1.60$ -- 3.45).
The value is also smaller than the result obtained for Hang Seng index
($\alpha =1.619$) \cite{wang2001}.
The value indicates that the central part of the IHSG distribution 
can be described by a L\'evy stable process. 

In the KLSE case the situation is rather different as depicted by 
the bottom panel of Fig.\,\ref{fig:distxyf0}. 
The PDF at zero return shows a cross-over at $\Delta t$ between 1 and 2
and, as a consequence, as shown by the solid line the 
linear regression to the points does not lead to a satisfactory result. 
To clarify this, we use the three first points in the second regression 
and excluding the $P(Z=0,\Delta t=1)$ data point in the third one. The 
results are compared with the linear regression to all points in 
the bottom-left panel of Fig.\,\ref{fig:distxyf0}. In the first case
we obtain $\alpha =1.7433$, already close to a Gaussian distribution
although still within the L\'evy stable index. In the second case the 
index $\alpha$ is smaller, indicating that the distribution for small
$\Delta t$ is far from Gaussian. However, in the third case the 
distribution is already a Gaussian unstable process. From this result,
clearly we can conclude that by slightly increasing $\Delta t$ the KLSE 
index does not retain its power law and quickly converges to a Gaussian
distribution, in contrast to the IHSG index. The PDF behavior for the 
three cases is clearly seen in the bottom-right panel of 
Fig.\,\ref{fig:distxyf0}.

The standard deviation $\sigma$ is known as the historical volatility
in financial literatures and quantifies the risk associated with 
the corresponding stock \cite{lillo2000}. As shown in the first line of Table 
\ref{tab:properties}, the KLSE stock is slightly riskier than 
the IHSG one. The kurtosis measures the relative peakedness of the 
distribution to a Gaussian one. The PDF of the less capitalized stocks 
is more leptokurtic than the PDF of the more capitalized ones 
\cite{lillo2000}. From Table \ref{tab:properties} we can clearly see 
that the IHSG stock is less capitalized than KLSE. 
Finally, the skewness characterizes the degree of
asymmetry of the distribution from its mean. A positive value of 
skewness indicates that the stock delivered more profits along its history,
whereas a negative value displays more losses hit the investor. Table
\ref{tab:properties} reveals that the IHSG stock in general 
gives more profits than the KLSE one. 

Although numerical values given in Table \ref{tab:properties} 
could be interesting, the values become much more important when
we discuss the behavior of the stock indices before and after the 
financial crash in the next section.

\begin{figure}[!t]
  \begin{center}
    \leavevmode
    \psfig{figure=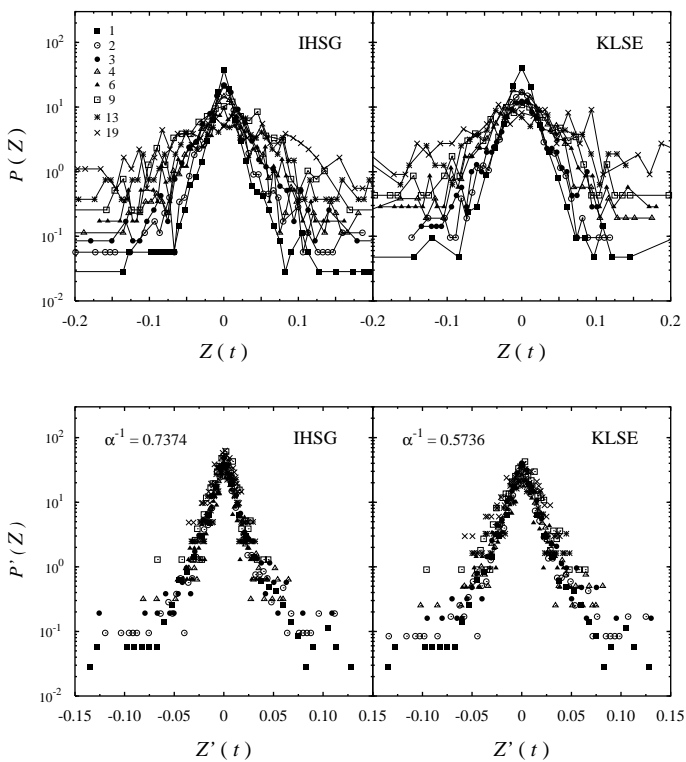,width=90mm}
    \caption{{\it Top panels}: Probability density functions of the IHSG
      and KLSE index returns for $\Delta t < 20$ days. 
      {\it Bottom panels}: 
      The same probability density functions, but plotted in scaled units,
      using Eqs.\,(\ref{eq:scaling1}) and (\ref{eq:scaling2}).
      Inverse of the L\'evy stable distribution index is shown in bottom 
      panels.}
   \label{fig:selfx} 
  \end{center}
\end{figure}

To further explore statistical properties of the PDF of index
returns we follow Refs.\,\cite{mantegna1995,mantegna1999}, i.e. we
investigate the stability of the distribution for $\Delta t > 1$ day.
Assuming the central part of the distribution can be described by 
a L\'evy stable distribution, then using the scaling variables 
\begin{eqnarray}
  \label{eq:scaling1}
  P'(Z) ~=~ P(Z) (\Delta t)^{1/\alpha} 
\end{eqnarray}
and
\begin{eqnarray}
  \label{eq:scaling2}
  Z'(t) ~=~ \frac{Z(t)}{(\Delta t)^{1/\alpha}} ~,
\end{eqnarray}
the empirical PDFs for different time sampling $\Delta t$ will
collapse onto the $\Delta t=1$ distribution.

The results for both indices are displayed in Fig.\,\ref{fig:selfx}, 
where we perform the analysis for $\Delta t < 20$ days. In general
the data collapse is evident, especially in the case of IHSG. In
the KLSE case we use $\alpha =1.7433$ (result from regression to all
empirical $P(0)$ in Fig.\,\ref{fig:distxyf0}), since the value leads to
an average behavior. As Ref.\,\cite{wang2001} has reported, we also
find that the extent of data collapse is stronger in the center of
the distribution. The existence of a cross-over in the KLSE case
is also observed here, the scaled $P'(Z)$ are more scattered than
that of the IHSG.

\subsection{Moments}
Reference \cite{gopi1999} has pointed out that the use of the return 
probability
to the origin ${\rm PDF}(0)$ to estimate the L\'evy stable distribution 
index $\alpha$ is statistically not optimal, due to discreteness of the
distribution. Instead of exploiting such method, Ref. \cite{gopi1999} 
used a different strategy, i.e. calculating  $\alpha$ by means of 
the slope of the cumulative distribution tails in a log-log plot.
To further test their results on the scaling behavior, 
Refs.\,\cite{gopi1999,plerou1999} analyzed the moments of the distribution
of normalized returns
\begin{eqnarray}
  \label{eq:moment}
  \mu (k) ~=~ \langle\, |g(t)|^k \rangle ~.
\end{eqnarray}
where the normalized returns $g(t)$ is defined by 
\begin{eqnarray}
  \label{eq:return_normalized}
  g(t) ~=~ \frac{Z(t)-\langle Z(t) \rangle}{\langle Z^2(t) \rangle
    - \langle Z(t) \rangle^2} ~,
\end{eqnarray}
with $\langle Z(t) \rangle$ the time average of $Z(t)$
over the entire of time series. In the case of the S\&P 500 index
the result is found to be consistent with the analysis of the tails of
cumulative distributions. They pointed out that the change in 
the moments behavior originates from the gradual disappearance of
the L\'evy slope in the distribution tails.
In our case it is also important to cross-check the results shown
in Fig.\,\ref{fig:distxyf0}, especially the KLSE slope, which is found
to be non-linear in the range of $\Delta t=1-10$.
 
\begin{figure}[tb]
  \begin{center}
    \leavevmode
    \psfig{figure=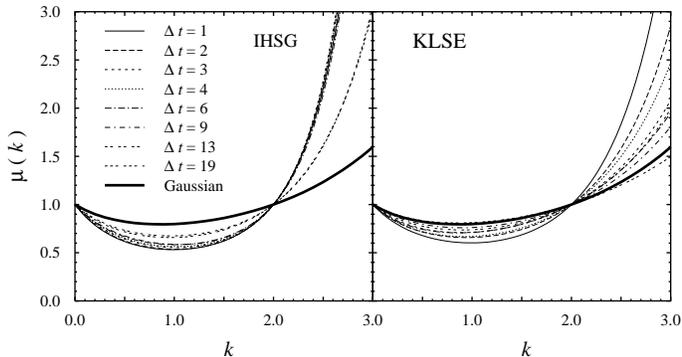,width=90mm}
    \caption{Moments of the distribution of normalized returns given in 
      Eq.\,(\ref{eq:moment}) for $\Delta t=1,2,3,4,6,9,13$, and 19 days
      compared with those obtained from the Gaussian distribution
      (solid thick lines). In the IHSG case, except for $\Delta t=13$
      and 19, all curves do not significantly deviate from the
      $\Delta t=1$ moment.}
   \label{fig:moment} 
  \end{center}
\end{figure}

It has been shown in Ref.\,\cite{gopi1999} that Eq.\,(\ref{eq:moment}) 
will diverge for $k\ge 3$. In this study we also constrain $k$ 
within $0\le k\le 3$. The result for both indices compared with 
the moment obtained from a Gaussian distribution are shown in
Fig.\,\ref{fig:moment}. Obviously the results are consistent with
our previous analysis, the IHSG moment retains its scaling up to 
$\Delta t=9$ days, only after $\Delta t=13$ the moment starts to 
deviate toward the Gaussian distribution. In the KLSE case the moment quickly
converges to the Gaussian distribution as $\Delta t$ increases from 1 day 
and does not show any scaling behavior as in the former case.

\subsection{Correlation in the stock index}
Another interesting behavior of the stock index fluctuation is its
short and long time memories which are usually expressed in terms of
the autocorrelation. Understanding the power law in the correlation
is very helpful in selecting the appropriate model or mechanism to
explain the fluctuation process.
In Ref.\,\cite{gopi1999} it has been shown that
the autocorrelation function of the S\&P 500 returns exhibits an exponential
decay with a characteristic decay time of approximately 4 minutes, whereas
the absolute value of index returns shows a scaling behavior with a
power-law exponent of 0.3. 
In our case it is of course difficult to study such behavior since both
indices are recorded daily. However, a qualitative comparison between
the two indices might help us to probe the differences 
in the two stocks.

\begin{figure}[tb]
  \begin{center}
    \leavevmode
    \psfig{figure=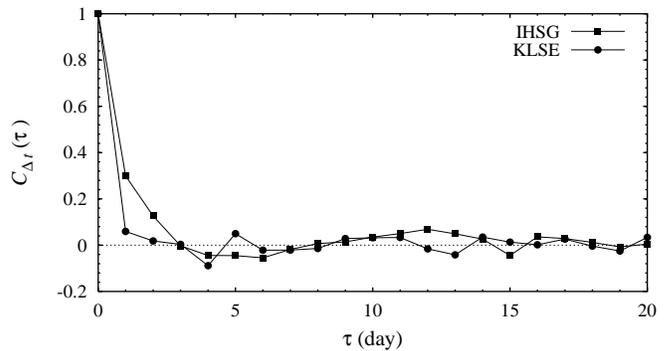,width=90mm}
    \caption{Short range autocorrelation functions of the IHSG and KLSE
      index returns sampled with $\Delta t=1$ day.}
   \label{fig:auto_short} 
  \end{center}
\end{figure}

Following Ref. \cite{gopi1999} the autocorrelation function is defined as
\begin{eqnarray}
  \label{eq:correlation}
  C_{\Delta t}(\tau) ~=~ \frac{\langle Z_{\Delta t}(t) Z_{\Delta t}(t+\tau) 
    \rangle - \langle Z_{\Delta t}(t)\rangle^2}{\langle Z_{\Delta t}^2(t)
    \rangle - \langle Z_{\Delta t}(t)\rangle^2} ~,
\end{eqnarray}
where $\tau$ indicates the time lag. The result for short-range 
autocorrelation is depicted in Fig.\,\ref{fig:auto_short}, where 
we can observe that the IHSG stock is slightly more correlated than the KLSE 
one. This result is certainly consistent with the analysis of the 
scaling and moments behavior. The fluctuation in the 
KLSE index seems to be more random than that of IHSG, as can be seen
also in Fig.\,\ref{fig:moment}.

\begin{figure}[tb]
  \begin{center}
    \leavevmode
    \psfig{figure=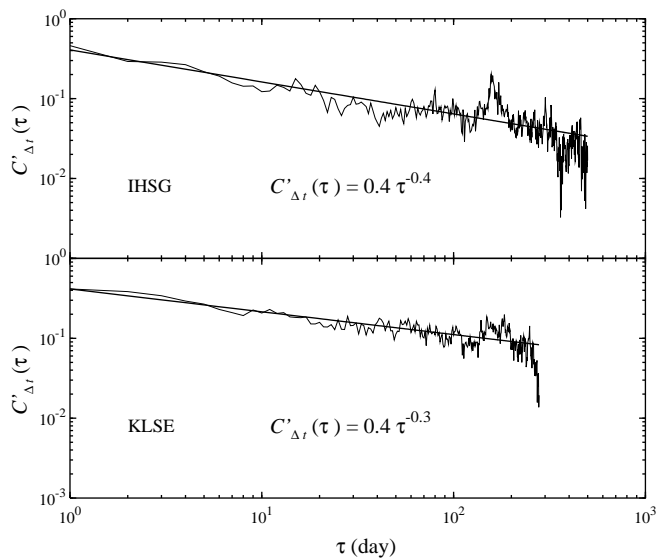,width=90mm}
    \caption{Log-log plots of the long-range autocorrelation function
      of the absolute index returns [see Eq.\,(\ref{eq:correlation})]. 
      Solid lines represent
      a linear regression over the entire range. Regression to the power law
      in the form of $a/(1+\tau^b)$ leads to a similar result.}
   \label{fig:auto_cor} 
  \end{center}
\end{figure}

The long-range autocorrelation function obtained by calculating the 
absolute value of the return in Eq.\,(\ref{eq:correlation}) 
is shown in Fig.\,\ref{fig:auto_cor}. 
The absolute value of index returns from both stocks show a long-range
power law 
behavior with a time scale up to almost one year. The only difference 
is observed in the power exponent of the scaling, the IHSG correlation 
falls off faster than the KLSE one. This is, however, in contrast to 
their short-range correlation. Note that Refs.\,\cite{liu1999,gopi1999}
fitted the long-range autocorrelation to the function in the
form of $a/(1+\tau^b)$ . In our analysis we found no significant difference
if we used such function.

\subsection{How volatile are the two stocks?}

\begin{figure}[!t]
  \begin{center}
    \leavevmode
    \psfig{figure=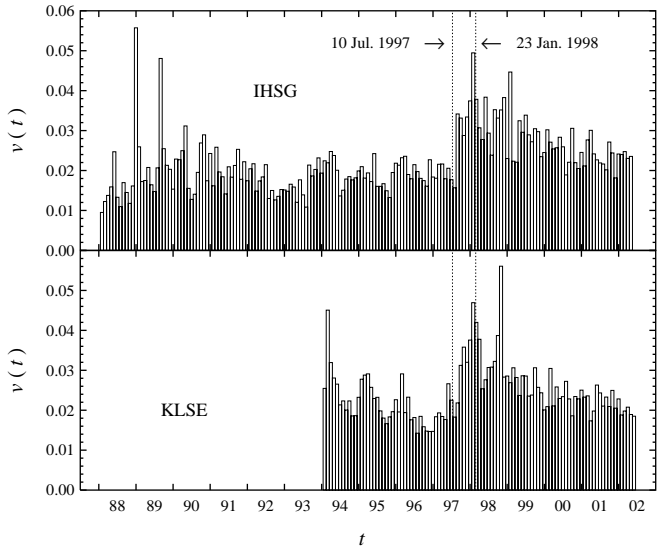,width=90mm}
    \caption{The volatility of the IHSG (top) and KLSE (bottom) stock
      indices obtained by calculating Eq.\,(\ref{eq:volatility})
      with $n=1$ month and sampling time $\Delta t=1$ day.}
   \label{fig:volat} 
  \end{center}
\end{figure}

Finally it is also interesting to see the volatility of the
two stocks. The volatility of the stock index is a measure of how much 
the index is likely to fluctuate. Information on the stock volatility
is very important for investors since it quantifies the risk factor
and it is also an important input of the Black-Scholes option-pricing
model. There are many different definitions of the volatility in 
financial literatures \cite{schwert}.
Here we define the volatility as an average of the absolute 
return over a time window with length $n$ sampled with $\Delta t=1$ day,
\begin{eqnarray}
  \label{eq:volatility}
  v(t) ~=~ \frac{1}{n} \sum_{\tau =t}^{t+n-1} \left| Z(\tau) \right| ~,
\end{eqnarray}
with $Z(\tau)$ is given by Eq.\,(\ref{eq:return}).

The results are shown in Fig. \ref{fig:volat}. As can be estimated from Figs.
\ref{fig:ihsg_return} and \ref{fig:klse_return} volatility in both stock 
indices increases after the financial crash. It is obvious that after the
drop of the foreign exchange rate finished in 23rd January 1998 
volatilities of both indices are still large until the indices reached 
their minimum values in September 1998. From that on,
the volatility decreases gradually although until 2002 the values are
still larger than the values before the crash.

It is also obvious from Fig.\,\ref{fig:volat} that before 1997 the IHSG
stock had a rather stable volatility variation, while on the other hand
the variation in the KLSE case was already dramatic. In general, the two
stocks maintain these behaviors after the crash, although in this period
the averaged volatilities are larger than before. This might explain why 
the IHSG stock index is found to be more correlated than the KLSE one.

\section{Statistical properties before and after crash}
\begin{table*}[!]
\renewcommand{\arraystretch}{1.2}
\caption{\label{tab:properties_bda} Statistical properties of IHSG and 
  KLSE stock indices before, during, and after crash.}
\begin{ruledtabular}
\begin{tabular}{lcccccc}
& \multicolumn{3}{c}{IHSG} & \multicolumn{3}{c}{KLSE} \\
\cline{2-4}\cline{5-7}
\raisebox{2.5ex}{Properties} & Before crash & During crash & After crash & 
Before crash & During crash & After crash \\
\hline
Average volatility  & 0.0192 & 0.0327 & 0.0260 & 0.0219 & 0.0341 & 0.0237 \\
Standard deviation $\sigma$ & 0.0167& 0.0300 & 0.0185 & 0.0127 & 0.0323 & 0.0197\\
Kurtosis & 183.421 & 2.849 & 4.377  & 6.775  & 8.152  & 42.925\\
Skewness & 6.8706 & 0.1417 & 0.7079 & 0.2053 & 1.0644 & $-0.2529$\\
$\gamma$ & $7.581\times 10^{-4}$ & $2.891\times 10^{-3}$ &
$9.723\times 10^{-6}$ & $9.339\times 10^{-6}$ & $3.109\times 10^{-4}$ &
$2.755\times 10^{-4}$ \\
$N$ & 2347 & 296 & 883 & 889 & 284 & 993 \\
\end{tabular}
\end{ruledtabular}
\end{table*}

\begin{figure}[!t]
  \begin{center}
    \leavevmode
    \psfig{figure=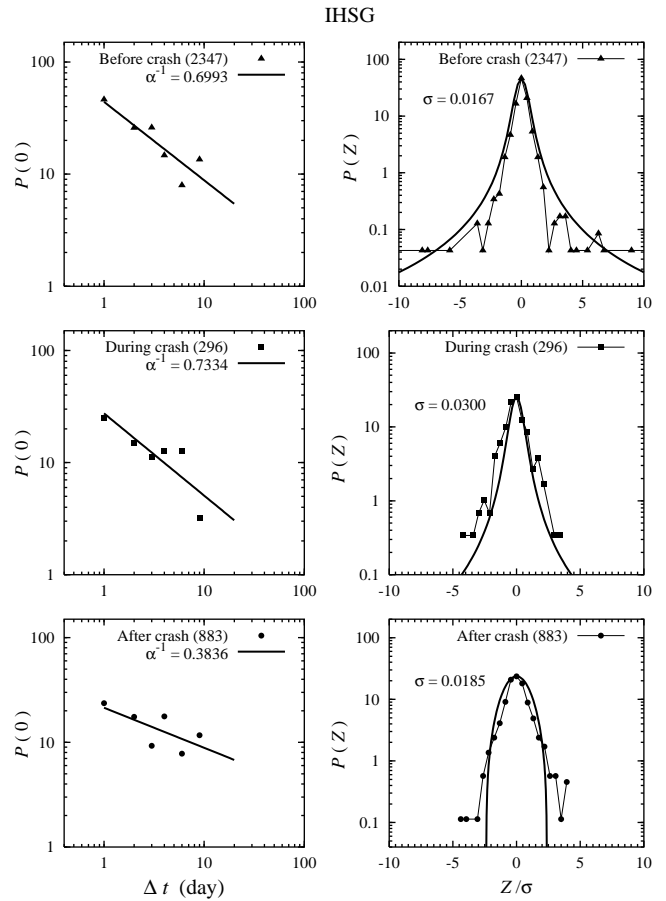,width=90mm}
    \caption{Same as in Fig.\,\ref{fig:distxyf0} but for the IHSG case 
      before, during, and after the financial crash.}
   \label{fig:distxyf0_bda} 
  \end{center}
\end{figure}

\begin{figure}[!]
  \begin{center}
    \leavevmode
    \psfig{figure=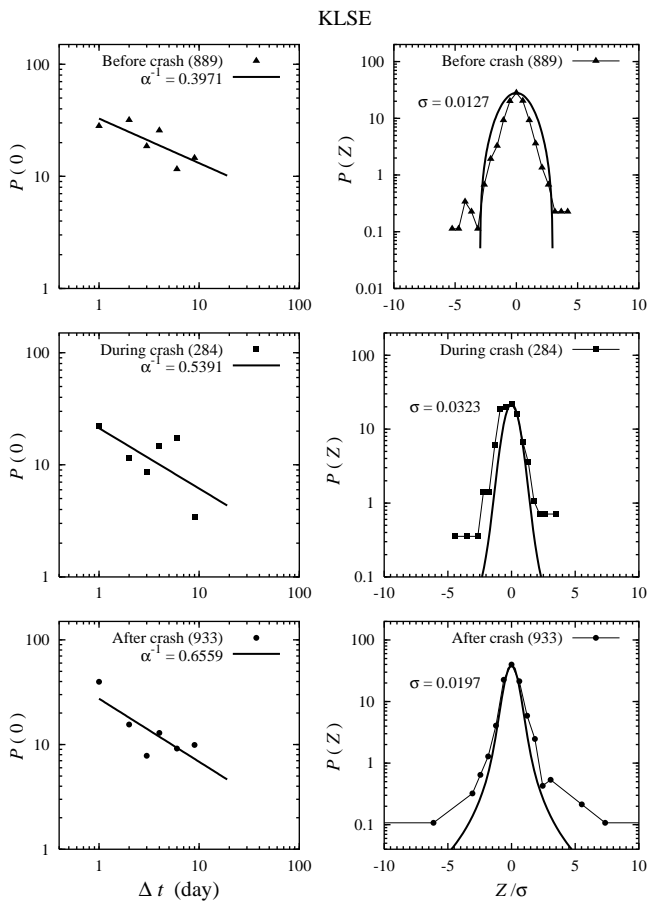,width=90mm}
    \caption{As in Fig.\,\protect\ref{fig:distxyf0_bda}, but for
      the KLSE stock index.}
   \label{fig:distxyf0_bda_klse} 
  \end{center}
\end{figure}

In section \ref{general} we have shown that between July 1997 and 
September 1998 the index returns of both stocks fluctuate strongly.
As shown in Fig.\,\ref{fig:volat}, compared with the situations
before and after this period, both indices become more volatile. 
A more quantitative description can be obtained from Table\,\ref{tab:properties_bda}, 
where in the first line we average the index volatility over all times before 
10th July 1997, 16 month after that (during which the volatility of
both indices is significantly large), and over the rest of data, respectively.
The result confirms the finding of Ref.\,\cite{lillo2000}, namely 
the volatility of a stock index tends to be large right after the crash
and in a relatively long period after that. This also happens to both stock
indices, i.e. almost five years after the financial crisis both stocks are still 
more volatile than before.

In the financial literature the volatility is often calculated from the
standard deviation of the returns distribution \cite{schwert}. 
In Table\,\ref{tab:properties_bda} we also present the standard deviations of
the IHSG and KLSE index returns distributions in all three cases. 
Table\,\ref{tab:properties_bda} indicates that both definitions are consistent
in our analysis.

From the kurtosis of stock indices we can see that the IHSG stock is more
capitalized after the crash, whereas the KLSE stock displays a very different
behavior. Nevertheless, the skewness of their distributions shows a similar tendency, 
both stocks deliver less profits after the crash, although more profits could 
be obtained from the KLSE stock during the crash.

\begin{figure}[!]
  \begin{center}
    \leavevmode
    \psfig{figure=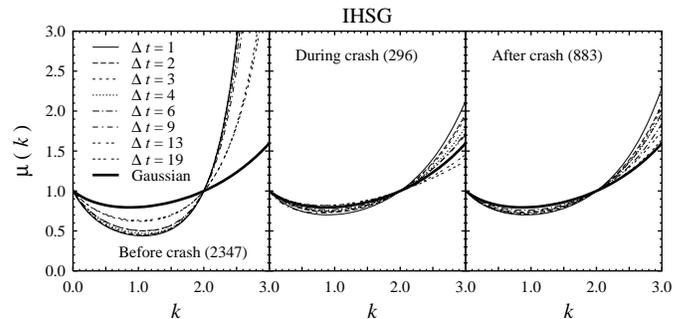,width=90mm}
    \caption{Moments of the distribution of IHSG normalized returns 
      before, during, and after crash.}
   \label{fig:momenti} 
  \end{center}
\end{figure}

\begin{figure}[!]
  \begin{center}
    \leavevmode
    \psfig{figure=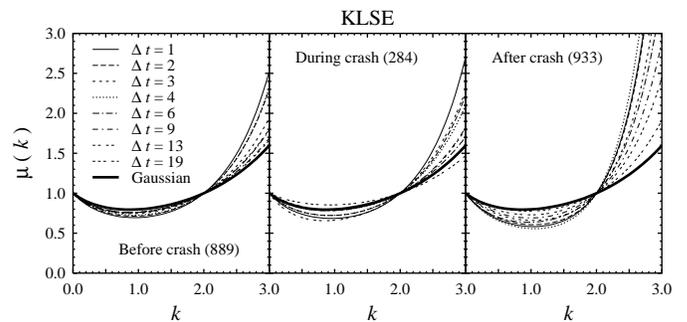,width=90mm}
    \caption{Same as in Fig.\,\ref{fig:momenti}, but for the KLSE index.}
   \label{fig:momentk} 
  \end{center}
\end{figure}

It is clearly tempting to investigate whether or not the universal properties
found in section \ref{general} still exist in all periods and to ask what kind
of differences might quantify these conditions. The answer 
can be found in Figs.\,\ref{fig:distxyf0_bda} and \ref{fig:distxyf0_bda_klse}.
Obviously in all cases the statistical accuracies are not as good as in the
previous section, when we used all data in our analyses. Nevertheless, in spite of 
the very limited data points used in our analysis, the message from 
Figs.\,\ref{fig:distxyf0_bda} and \ref{fig:distxyf0_bda_klse} is clear,
the universal properties of the returns distribution 
retain their existences in all cases. Linear regression 
to the IHSG probability of the return to the origin reveals 
the fact that the distribution
turns from L\'evy to Gaussian after the crash. Surprisingly, the KLSE stock 
index shows a contrary result, the distribution alters from a Gaussian to 
a L\'evy one after the crash.

A cross-check to the result shown in Figs.\,\ref{fig:distxyf0_bda} 
and \ref{fig:distxyf0_bda_klse} is inevitable, since the number of 
data could significantly limit statistical accuracies in this case.
For this purpose in Figs.\,\ref{fig:momenti} and \ref{fig:momentk} we
display the moments of both stock indices in the case of before, during,
and after the financial crash. A consistent result is obtained from
these figures, in the case of IHSG the distribution becomes closer
to Gaussian during and after crash, whereas the KLSE moments move away
from the Gaussian distribution after the crash.

Another important finding obtained from Fig.\,\ref{fig:momenti} is that the
scaling behavior up to $\Delta t=9$ shown by the IHSG case in the 
previous section originates from the period before crash. After the crash,
the IHSG moments quickly converge to a Gaussian distribution. In fact,
this phenomenon has already been seen in the left panels of 
Fig.\,\ref{fig:distxyf0_bda}, where the empirical 
probability of return to the origin is more scattered in the crash periods 
and after that.

\begin{figure}[!]
  \begin{center}
    \leavevmode
    \psfig{figure=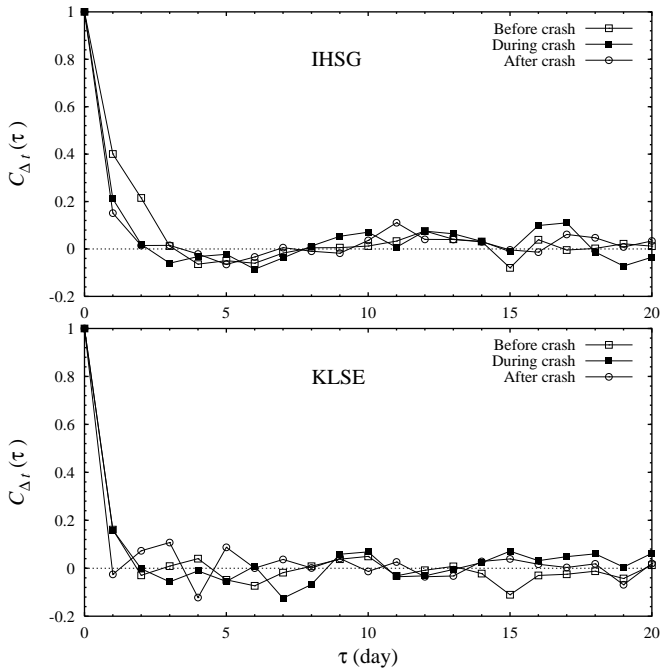,width=90mm}
    \caption{Short range autocorrelation functions of the IHSG and KLSE
      index returns before, during, and after crash.}
   \label{fig:auto_short_bda} 
  \end{center}
\end{figure}

As in the previous section,
investigation of the autocorrelation function from both stocks 
might also be of interest. The corresponding autocorrelation is shown 
in Fig.\,\ref{fig:auto_short_bda}.
Although it is difficult to see how long the KLSE stock maintains its 
correlation, the IHSG stock index obviously becomes less correlated 
during and after the crash. This result can be understood by looking
back to Fig.\,\ref{fig:momenti}, i.e. after the crash the returns distribution 
becomes more random and therefore the index losses its memory.

It might be more interesting if we compare results in this section 
with currency controls shown in the top panels of 
Figs.\,\ref{fig:ihsg_return} and \ref{fig:klse_return}.
The government interventions on the exchange rate seem to create 
non natural economics activities which have a direct influence on
the fluctuation of stock indices. Once the currency is under-controlled,
the fluctuations tend to significantly deviate from the Gaussian distribution,
which can be clearly seen in the IHSG moment before crash (Fig.\,\ref{fig:momenti})
and the KLSE one after crash (Fig.\,\ref{fig:momentk}). From the investors
point of view this is obvious, a less fluctuating exchange rate reduces 
the number of variables considered in estimating the future stock price.

\section{Conclusions}
We have analyzed the IHSG and KLSE stock index returns using the methods
developed for statistical physics. In spite of the limited number of data
used in our analysis, we still found that both stock indices show 
the universal properties previously observed in the leading stocks such as S\&P 500 
and NYSE, i.e. the scaling properties. The difference is, nevertheless, 
found in the time
scale, the IHSG stock retains its scaling property longer than the KLSE one.
As a consequence, the former is more correlated than the latter.

By investigating their statistical properties before, during, and after the
financial crash it is found that the scaling behavior of the IHSG index
originates from its fluctuation before the crash period. During and after
the crash periods the index distributions are closer to Gaussian and do not
show a stable process. In the KLSE case, the return distributions in all 
periods show an unstable process and depart from Gaussian after the crash. 
One possible explanation to these different behaviors could be the currency 
control. Besides these differences, both stock indices 
show some similarities, namely both stocks become more volatile during and
after the crash and their return moments become closer to Gaussian during
the crash. 

\acknowledgments
This work was supported in part by the Quality for Undergraduate
Education (QUE) project. The author thanks Dwi Arsono for collecting the
IHSG data and to Dr. Chairul Bahri for careful reading of the manuscript
and critical comments.

\end{document}